\documentclass[aps,prl,showpacs,twocolumn,superscriptaddress]{revtex4-2}
\usepackage[pdftex]{graphicx} %[dvips]
\usepackage{natbib}
\usepackage{booktabs}

\usepackage[colorlinks, citecolor=blue, linkcolor=blue, urlcolor=blue, breaklinks]{hyperref}
\usepackage[svgnames]{xcolor}
\usepackage{changes}

\newcommand{\SMO}{SrMoO$_3$}
\newcommand{\SRO}{SrRuO$_3$}

\begin{document}
\title{Electronic structure of the highly conductive perovskite oxide \SMO}

\author{E. Cappelli}
\affiliation{Department of Quantum Matter Physics, University of Geneva, 24 Quai Ernest-Ansermet, 1211 Geneva 4, Switzerland}
\author{A. Hampel}
\affiliation{Center for Computational Quantum Physics, Flatiron Institute, 162 Fifth Avenue, New York, NY 10010, USA}

\author{A. Chikina}
\affiliation{Swiss Light Source, Paul Scherrer Institut, CH-5232 Villigen PSI, Switzerland}
\author{E. Bonini Guedes}
\affiliation{Swiss Light Source, Paul Scherrer Institut, CH-5232 Villigen PSI, Switzerland}
\author{G. Gatti}
\affiliation{Department of Quantum Matter Physics, University of Geneva, 24 Quai Ernest-Ansermet, 1211 Geneva 4, Switzerland}
\author{A. Hunter}
\affiliation{Department of Quantum Matter Physics, University of Geneva, 24 Quai Ernest-Ansermet, 1211 Geneva 4, Switzerland}
\author{J. Issing}
\affiliation{Department of Quantum Matter Physics, University of Geneva, 24 Quai Ernest-Ansermet, 1211 Geneva 4, Switzerland}
\author{N. Biskup}
\affiliation{GFMC, Departamento de F\'{i}sica de Materiales, Universidad Complutense de Madrid, 28040 Madrid, Spain}
\affiliation{Instituto Pluridisciplinar, Universidad Complutense de Madrid, 28040 Madrid, Spain}
\author{M. Varela}
\affiliation{GFMC, Departamento de F\'{i}sica de Materiales, Universidad Complutense de Madrid, 28040 Madrid, Spain}
\affiliation{Instituto Pluridisciplinar, Universidad Complutense de Madrid, 28040 Madrid, Spain}
\author{Cyrus E. Dreyer}
\affiliation{Department of Physics and Astronomy, Stony Brook University, Stony Brook, New York, 11794-3800, USA}
\affiliation{Center for Computational Quantum Physics, Flatiron Institute, 162 Fifth Avenue, New York, NY 10010, USA}
\author{A. Tamai}
\affiliation{Department of Quantum Matter Physics, University of Geneva, 24 Quai Ernest-Ansermet, 1211 Geneva 4, Switzerland}
\author{A. Georges}
\affiliation{Center for Computational Quantum Physics, Flatiron Institute, 162 Fifth Avenue, New York, NY 10010, USA}
\affiliation{Coll\`{e}ge de France, 11 Place Marcelin Berthelot, 75005 Paris, France}
\affiliation{PHT, CNRS, \'{E}cole Polytechnique, IP Paris, F-91128 Palaiseau, France}
\affiliation{Department of Quantum Matter Physics, University of Geneva, 24 Quai Ernest-Ansermet, 1211 Geneva 4, Switzerland}
\author{F.Y. Bruno}
\affiliation{GFMC, Departamento de F\'{i}sica de Materiales, Universidad Complutense de Madrid, 28040 Madrid, Spain}
\author{M. Radovic}
\affiliation{Swiss Light Source, Paul Scherrer Institut, CH-5232 Villigen PSI, Switzerland}
\author{F. Baumberger}
\affiliation{Department of Quantum Matter Physics, University of Geneva, 24 Quai Ernest-Ansermet, 1211 Geneva 4, Switzerland}
\affiliation{Swiss Light Source, Paul Scherrer Institut, CH-5232 Villigen PSI, Switzerland}

\date{\today}

\begin{abstract}
We use angle-resolved photoemission to map the Fermi surface and quasiparticle dispersion of bulk-like thin films of \SMO{} grown by pulsed laser deposition. The electronic self-energy deduced from our data reveals weak to moderate correlations in \SMO, consistent with our observation of well-defined electronic states over the entire occupied band width.
We further introduce spectral function calculations that combine dynamical mean-field theory with an unfolding procedure of density functional calculations and demonstrate good agreement of this approach with our experiments.
\end{abstract}

\maketitle

% INTRODUCTION
%
% High conductivity
\SMO{} stands out among perovskite transition metal oxides because of its exceptionally low room-temperature resistivity of $\approx5.1$~$\mu\Omega \, \text{cm}$, only about three times that of copper~\cite{nagai2005}. This remarkably high conductivity has sparked an interest into possible applications of \SMO{} in microwave electronics, as plasmonic devices or as electrodes in oxide heterostructures~\cite{radetinac2017, salg2019, walk2019, salg2020, stoner2019, wells2017, wells2018b}.
The high conductivity of \SMO{} is particularly remarkable when placed in the context of other 4$d$ perovskite transition metal oxides. The Mo$^{4+}$ ion in \SMO{} has nominally 2 electrons in the $4d\:t_{2g}$ shell and is thus particle-hole symmetric to the $4d^4$ configuration of the Ru$^{4+}$ ion in \SRO, 
although of course the full electronic structure of these two materials is not related by this symmetry.
Yet, \SRO{} has a room-temperature resistivity of $\approx 200$~$\mu\Omega \, \text{cm}$, approximately 40 times higher than \SMO~\cite{kikugawa2015}. Density functional theory plus dynamical mean field theory (DFT+DMFT) calculations further found that the resisitvity of \SRO{} at elevated temperature is dominated by electron-electron interactions~\cite{deng2016}. Together, this suggests unusually weak electronic correlations in \SMO, consistent with its low Sommerfeld coefficient of \mbox{$\gamma \approx 7.9$~mJ/(mol$\,$K$^2$)}~\cite{ikeda2000, nagai2005, zhao2007, wadati2014, karp2020} as compared to \mbox{$\gamma \approx 30$~mJ/(mol$\,$K$^2$)} for \SRO~\cite{kikugawa2015}. However, to date, little is known from experiment about the electronic self-energy and quasiparticle effective mass of \SMO.

% Growth by others
The growth of bulk single crystals of \SMO{} proved exceptionally difficult because the Mo$^{4+}$ oxidation state is only stable under strongly reducing conditions~\cite{kamata1975, nagai2005}. Nagai~\textit{et al.} reported that at the melting point of \SMO{} of $\approx 2000$~K, oxygen pressures no higher than $ 10^{-22}$~mbar are required to suppress the more stable SrMoO$_4$ phase and stabilize the Mo$^{4+}$ ion of \SMO{}~\cite{nagai2005}. Many studies thus focused on powder samples or epitaxial thin films which can be stabilized at significantly lower temperatures where requirements on the oxygen partial pressure are more relaxed~\cite{brixner1960, wang2001b}. High-quality epitaxial thin films of \SMO{} have been successfully grown by pulsed-laser deposition (PLD) on SrTiO$_3$ and GdScO$_3$ substrates, with the most conductive ones reaching a room-temperature resistivity of 20~$\mu\Omega \, \text{cm}$~\cite{alff2014}. Recently, thin films with similar resistivity were also obtained by molecular beam epitaxy on KTaO$_3$ substrates with an SrTiO$_3$ buffer layer and were characterized by soft X-ray angle-resolved photoelectron spectroscopy (ARPES)~\cite{takatsu2020}.

% Crystalline and electronic structure
Neutron diffraction on polycrystalline \SMO{} samples revealed two structural phase transitions. At room temperature, bulk \SMO{} is an ideal cubic $Pm\bar{3}m$ perovskite. When the temperature is lowered, it first transitions to a tetragonal $I4/mcm$ phase below $\approx 250$~K and then to an orthorhombic $Imma$ ground state below $\approx 150$~K~\cite{macquart2010}. The nominal valence Mo$^{4+}$ has been confirmed by X-ray photoelectron spectroscopy (XPS)~\cite{wadati2014} and is consistent with Hall-effect data~\cite{wang2001b, wadati2014} showing close to the expected 2 electrons per molybdenum site. To date, no signs of superconductivity or magnetic ordering were detected in \SMO{}~\cite{bouchard1968, ikeda2000} although a recent theoretical study suggests proximity to an antiferromagnetic state at very low temperatures~\cite{hampel2021}.

\begin{figure*}[tb]
  \includegraphics[width=0.8\textwidth]{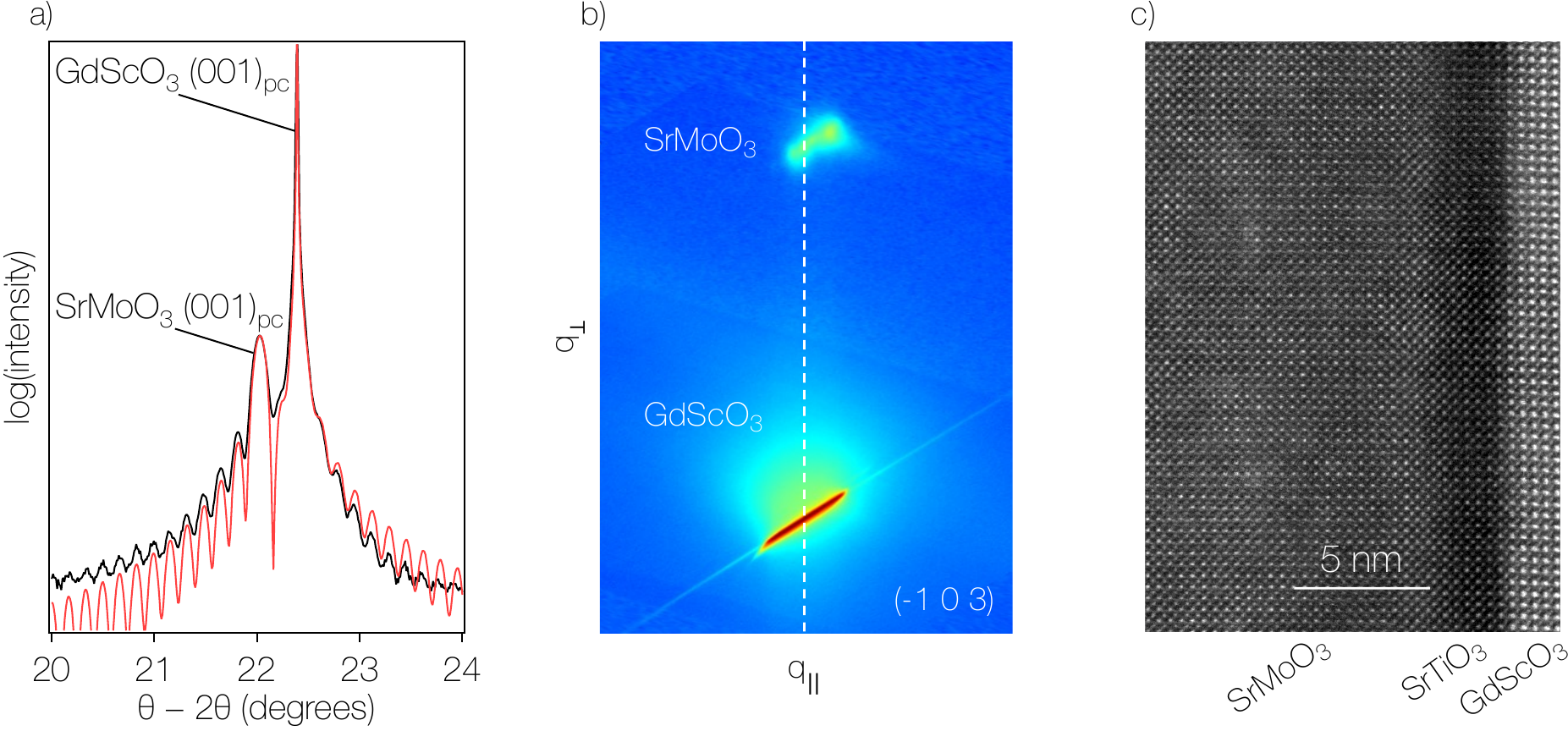}
  \caption{Characterization of the \SMO{} thin films. a) XRD $\theta - 2\theta$ scan around the pseudo-cubic (001) Bragg reflection (black) and the corresponding fit (red), performed with the \textit{InteractiveXRDFit} software~\cite{lichtensteiger2018}. b) Reciprocal-space map around the ($\bar{1}03$) GdScO$_3$ peak, shown on a logarithmic color scale. The vertical dashed line serves as a guide to the eye. c) HAADF-STEM image of a \SMO{} film with a 5-unit-cell SrTiO$_3$ buffer layer. 
In high angle annular dark field (HAADF) mode, also known as $Z$-contrast, the contrast of every atomic column is roughly proportional to the atomic number Z squared.
 The data in (b) and (c) were acquired on the same film whereas the XRD profile in (a) was obtained on a \SMO{} film without SrTiO$_3$ buffer layer.}
  \label{main1}
\end{figure*}

Here, we present a comprehensive ARPES data set on \SMO{} thin films grown on GdScO$_3$ substrates. 
We reveal the effect of orthorhombic distortions on the electronic structure and quantitatively analyze the electronic self energy of \SMO. We further show that a DFT+DMFT calculation provides an accurate description of our data.

%
% EXPERIMENTAL METHODS
%

\begin{figure*}[tb]
  \includegraphics[width=0.85\textwidth]{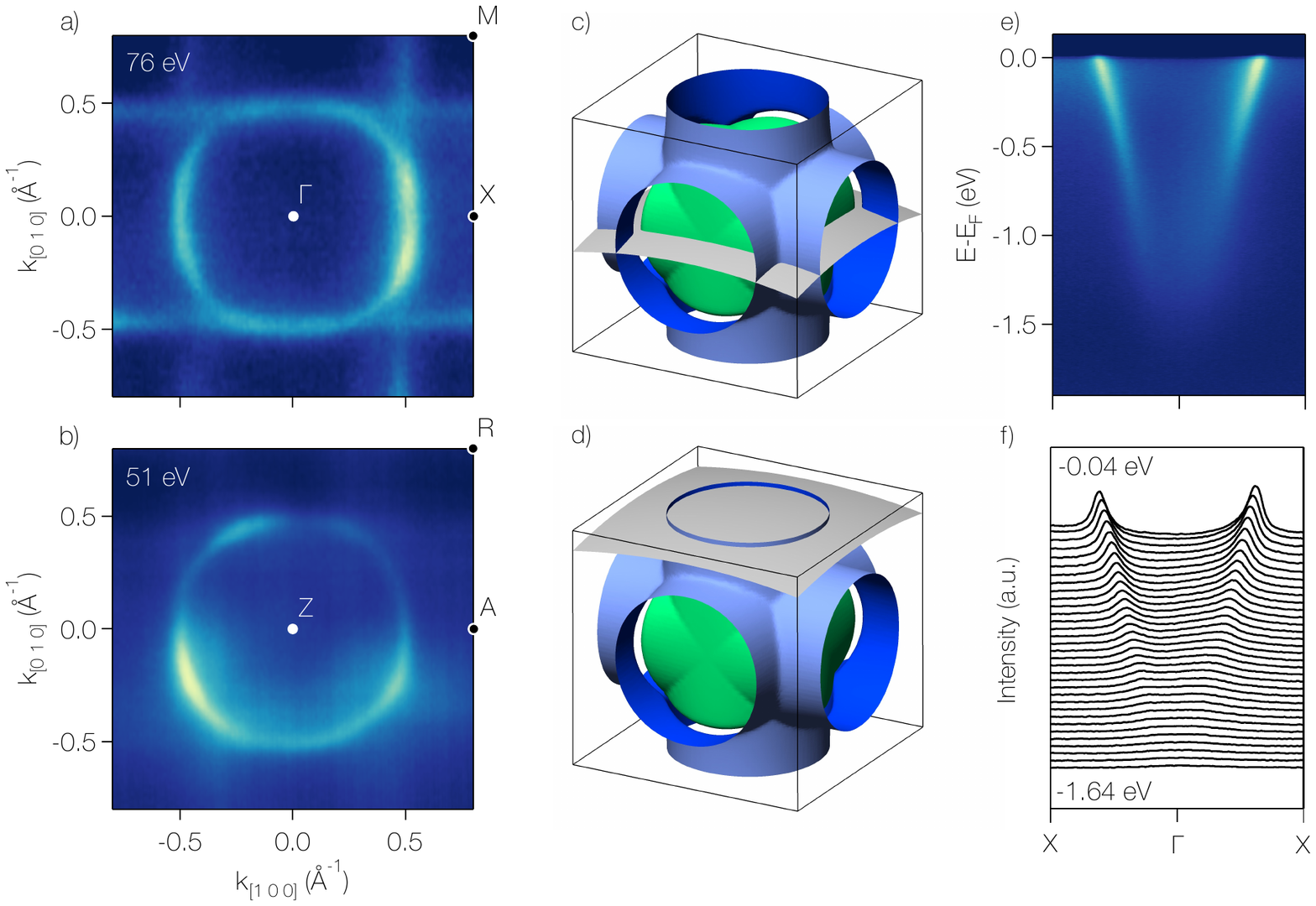}
  \caption{Salient electronic structure of bulk-like \SMO{} films. a, b) Experimental Fermi-surface maps measured at the center and boundary of the three-dimensional Brillouin zone, using circularly-polarized photons with energy 76~eV and 51~eV, respectively. c, d) DFT Fermi surface of cubic $Pm\bar{3}m$ \SMO{}. The gray spherical surfaces represent the nearly free-electron photoelectron final states for the photon energies used in (a,b). e) Energy-momentum measured in the second Brillouin zone with a photon energy of 86~eV to probe the X$\Gamma$X line. f) Stack of MDCs extracted from the data in (e).}
  \label{main2}
\end{figure*}

% Growth by us
We prepared highly crystalline epitaxial thin films of \SMO{}, both by sputter deposition at the University of Geneva and by PLD at the Swiss Light Source (SLS)~\cite{cappelli2021}. In order to access the full three-dimensional electronic structure, we performed the ARPES measurements shown in the main text with synchrotron radiation at the SLS. In the following, we therefore focus on the PLD-grown films. Additional ARPES data on sputtered films is shown in Ref.~\cite{cappelli2021} and supplementary information section IV. 
The PLD films were grown on orthorhombic GdScO$_3$ substrates in vacuum ($p\approx 10^{-7}$~mbar), so as to avoid the formation of the fully oxidized SrMoO$_4$ phase. 
The samples were transferred to the SIS ULTRA end station using either a vacuum suitcase or an ultra-high vacuum transfer system. ARPES experiments at SLS were performed at a temperature of $\approx10$~K with circularly polarized light and photon energies ranging from 40 to 200~eV. The energy resolution of the ARPES experiments changed with photon energy and was in the range of $20-40$~meV for the data shown in the main text.
Some of the films were grown on top of a SrTiO$_3$ buffer layer of $5 - 10$ unit cells (u.c.). While we found that the buffer promotes layer-by-layer growth during the early stages of the deposition, no significant difference was observed in films of 10 or more unit cells either by ARPES, x-ray diffraction (XRD) or reflection high-energy electron diffraction (RHEED). The best results were obtained for substrate temperatures between $800$ and $900^{\circ}$C, as measured by a pyrometer.
Electron microscopy measurements were carried out in a JEOL ARM200cF equipped with a Ceoss aberration corrector and a Gatan Quantum energy filter. Specimens were prepared by conventional methods: mechanical grinding and Ar ion milling.

Density functional theory (DFT) calculations were performed with the Vienna Ab initio Simulation Package (VASP)~\cite{Kresse:1993bz,Kresse:1996kl,Kresse:1999dk}, with the exchange-correlation functional of Perdew, Burke, and Ernzerhof~\cite{Perdew:1996iq}. The DFT calculations use a dense $41 \times 41 \times 41$ k-point mesh for the cubic unit cell (same density for the orthorhombic cell) to avoid $k$-point discretization issues.

Dynamical mean field theory (DMFT) calculations are performed using solid\_dmft~\cite{solid_dmft} software, which utilizes the \textsc{TRIQS/DFTTools} software package~\cite{aichhorn_dfttools_2016,parcollet_triqs_2015}. 
For these calculations, we construct a realistic low-energy Hamiltonian containing only the three Mo $t_{2g}$ like states around the Fermi level using maximally localized Wannier functions as implemented in \textsc{wannier90}~\cite{Pizzi_2020}. 
The effective impurity problem within the DMFT cycle is solved with the numerically exact real frequency fork tensor product state (FTPS) solver~\cite{Bauernfeind:2017}. The calculations were well converged down to a bath discretization broadening $\eta=0.1$~eV without showing finite size effects in the real time Green function. We add a local Coulomb interaction in the form of the Hubbard-Kanamori Hamiltonian including all spin-flip and pair-hopping terms~\cite{vaugier2012} with parameters $U=3.40$~eV and $J=0.33$~eV, close to the values obtained from cRPA calculations in Ref.~\cite{hampel2021}.

% Sample characterization -- emphasis on orthorhombic experimental unit cell
The structural quality of the samples was characterized both by \textit{in-situ} RHEED and by \textit{ex-situ} XRD. The XRD scan in Fig.~\ref{main1} (a) shows a \SMO{} peak with pronounced finite-thickness oscillations, which attest to the high crystalline coherence of the film. By fitting the experimental data we extract a film thickness $d \approx 50$~nm and a pseudo-cubic out-of-plane lattice constant $c_{pc} = 4.030 \pm 0.002$~\r{A}~\cite{lichtensteiger2018}. The vertical alignment of film and substrate peaks in the reciprocal-space map in Fig.~\ref{main1}~(b) demonstrates that the \SMO{} thin films are fully strained to the GdScO$_3$ substrates, as confirmed by high resolution scanning electron microscopy (STEM) images such as the one in Fig.~1(c). 
Our \SMO{} thin films are thus orthorhombic, which implies the presence of octahedral tilts and therefore a doubling of the c-axis. 
Hence we deduce an orthorhombic unit cell with lattice parameters $(a_o, b_o, c_o) \equiv (\sqrt{2} a_{pc}, \sqrt{2} b_{pc}, 2 c_{pc}) = (5.49, 5.75, 8.06)$~\r{A}, where $(a_o, b_o)$ are the in-plane lattice parameters of the GdScO$_3$ substrate at room temperature.

% Sample characterization -- resistivity
Resistivity measurements were performed in a Physical Property Measurement System (PPMS), following the standard van der Pauw procedure~\cite{vdpauw1958, nist} with platinum contacts deposited at the four corners of the samples. Our films display metallic behavior at all temperatures, with a room-temperature resistivity between 60 and 70~$\mu\Omega \, \text{cm}$ and a residual-resistivity ratio (RRR) of $\approx 1.5$. The resistivity of our samples is comparable to previously reported data from \SMO{} thin films grown by PLD~\cite{wang2001a, lekshmi2005}, although the best-quality films reported in the literature are about $2 - 3$ times more conductive~\cite{alff2014, radetinac2016, takatsu2020}. The resistivity of all films closely follows the Fermi liquid form $\rho (T) = \rho_0 + A T^2$, with 
$A = 3.8(1) \cdot 10^{-10}$~$\Omega \, \text{cm} / \text{K}^2$ up to $\approx 100$~K. 
Further details regarding the growth and characterization can be found in the Supplementary Information.

%
% RESULTS AND DISCUSSION
%

\begin{figure*}[tb]
  \includegraphics[width=0.8\textwidth]{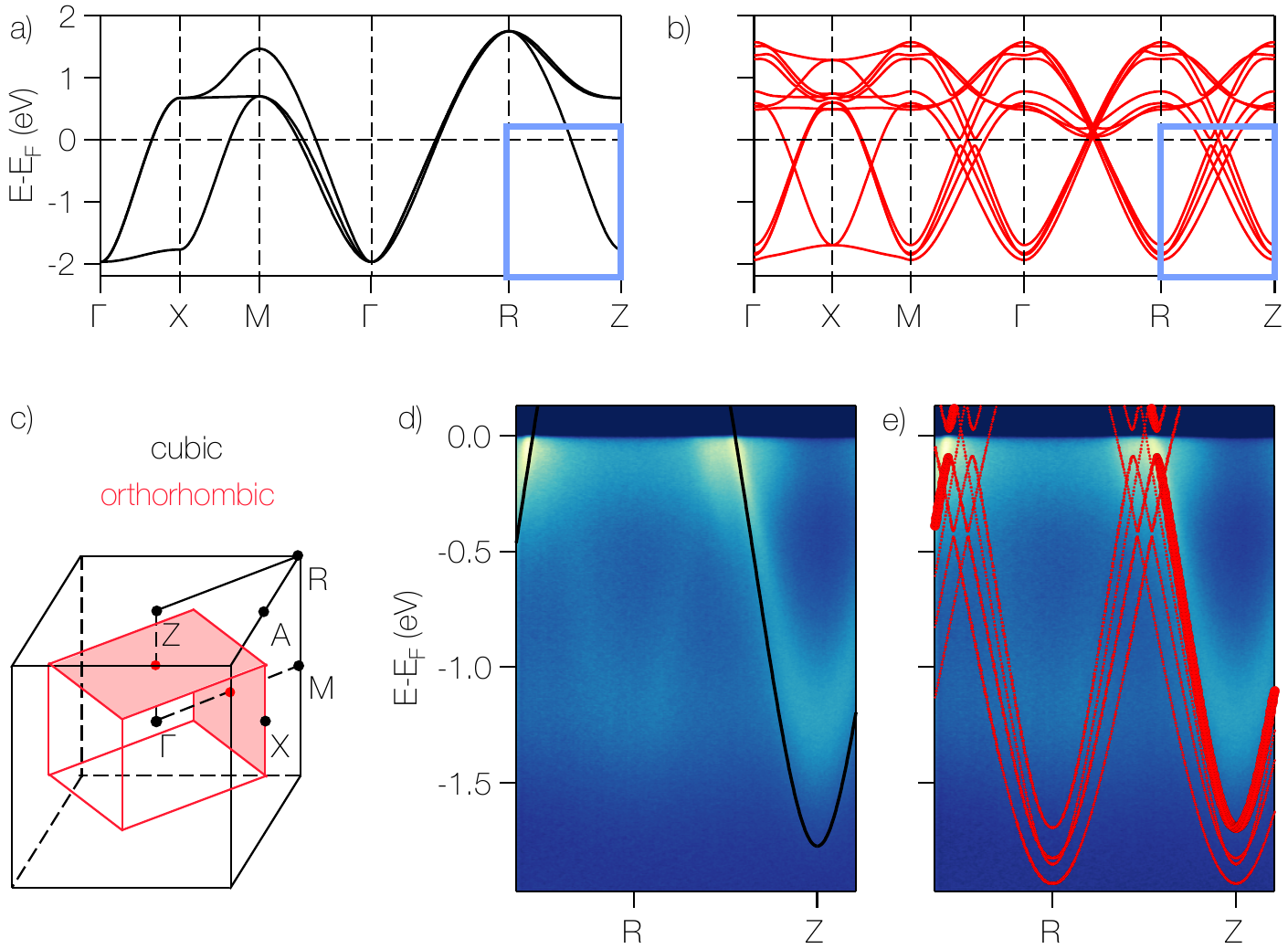}
  \caption{a,b) Band dispersion for the ideal simple-cubic unit cell and the experimental orthorhombic unit cell, respectively. The light blue box highlights the momentum space region of the ARPES data shown in (d,e). c) Sketch of the first Brillouin zone for a simple-cubic lattice (black) and for an orthorhombic supercell of the same lattice (red). d) ARPES energy-momentum cut at $h\nu=60$~eV, probing the electronic structure close to the ZR high-symmetry line.
The DFT simple-cubic band dispersion along ZR is overlaid in black. e) Same data with the DFT band dispersion obtained for the relaxed experimental orthorhombic unit cell in red. The thickness of the lines represents the unfolded spectral weights.}
  \label{main3}
\end{figure*}

\begin{figure*}[tb]
  \includegraphics[width=0.85\textwidth]{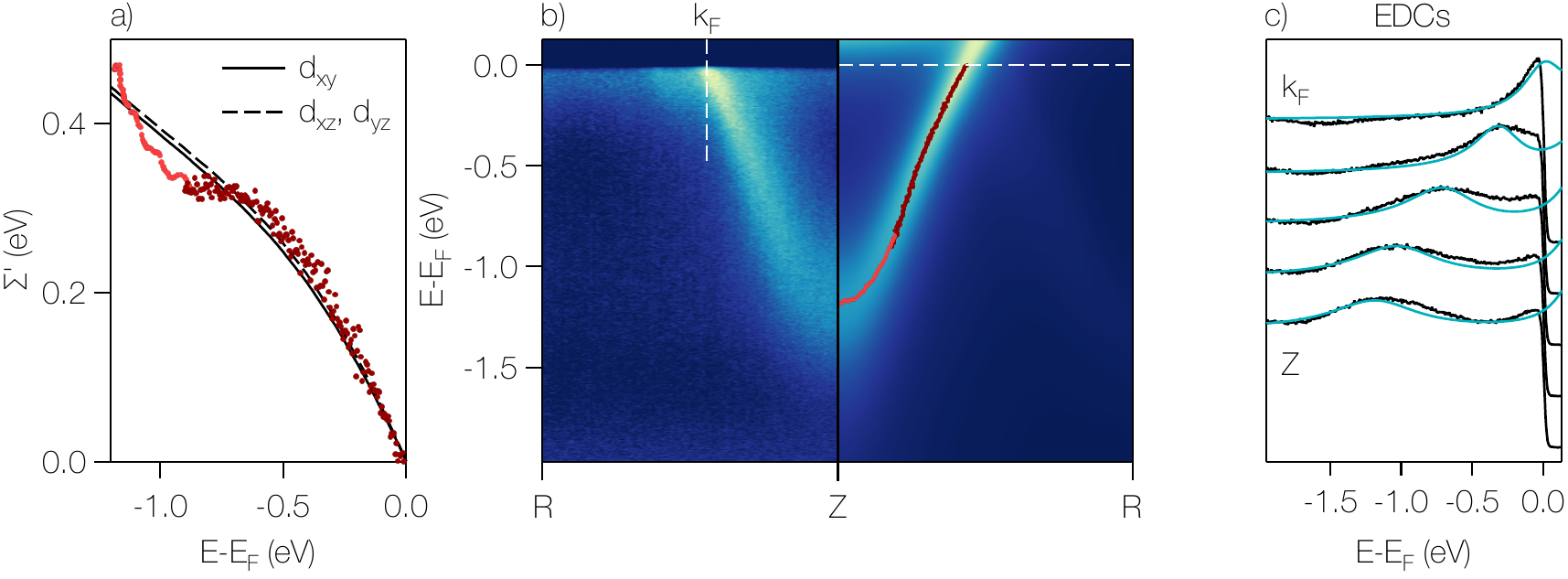}
  \caption{Electronic self-energy and comparison with DMFT calculations. a) Real part of the electronic self-energy $\Sigma'$ extracted from the band dispersion along the ZR high-symmetry direction (dark red for fits to MDCs, light red for fits to EDCs), together with the corresponding DMFT calculation for the experimental orthorhombic unit cell (black solid and dashed). b) \textit{Left:} ARPES energy-momentum cut at $h\nu=70$~eV along the ZR high-symmetry direction. \textit{Right}: DMFT spectral function along the mirrored path. The quasiparticle dispersion extracted from fits to MDCs and EDCs extracted from the experimental data is overlaid in dark and light red, respectively. c) Selected experimental EDCs from the data in Fig.~4(b) together with the corresponding EDCs from DMFT (turquoise). $k_{F}$ is marked by a vertical dashed line in panel (b).}
  \label{main4}
\end{figure*}

% The 3D electronic structure (ideal cubic)
Figure~\ref{main2} illustrates the overall electronic structure of bulk \SMO{} in the pseudo-cubic Brillouin zone containing one Mo site. The Fermi surface in panel (a) -- measured at a photon energy corresponding to the bulk $\Gamma$ point -- shows a circular electron pocket centered at $\Gamma$ and two orthogonal sets of straight contours tangential to it. Assuming a cubic symmetry of the dominant ionic potential terms, this implies a bulk Fermi surface with 3 interpenetrating orthogonal cylinders, in good agreement with the DFT calculation for cubic \SMO{} shown in (c,d). 
Changing the photon energy to probe the electronic structure near the boundary of the pseudo-cubic Brillouin zone, we thus expect a single, circular Fermi surface contour centered at the $Z$ point. This is directly confirmed in Fig.~\ref{main2}~b).
A quantitative comparison of experimental and theoretical Fermi surface contours as well as a Fermi surface in the $k_z$-plane ($k_{[100]}-k_{[001]}$) are shown in Supplementary Information.

The 3 quasi-2D Fermi surface sheets of \SMO{} can readily be identified with the 3 $t_{2g}$ orbitals, which each disperse strongly along 2 cubic axes and weakly along the third one. The hybridization gaps at the intersection of the 3 cylindrical Fermi surface sheets appear to be small and cannot be resolved directly in our experiment.
The energy-momentum cut along the X$\Gamma$X high-symmetry direction shown in Fig.~\ref{main2}~e) confirms the electron-like nature of the Fermi surface cylinders. Remarkably, the quasiparticle band can be traced over the entire occupied band width of $\approx 1.2$~eV. This is in stark contrast to other transition metal oxides such as ruthenates where electronic states remain well-defined and coherent near the Fermi surface only.

% Cubic vs orthorhombic - qualitative discussion
In Figure~\ref{main3} we discuss the modification of the electronic structure of \SMO{} due to the orthorhombic distortions. Our X-ray diffraction measurements imply an orthorhombic primitive unit cell with 4 Mo ions. Hence, we expect 12 bands derived form the Mo $t_{2g}$ manifold. To a first approximation, these bands can be described by back-folding the simple cubic electronic structure into the orthorhombic Brillouin zone. This effect is clearly visible in the DFT bands for the two unit cells shown in Fig.~\ref{main3}(a,b). The backfolding is also evident in the ARPES data in Fig.~\ref{main3}(d,e) in the form of a parabolic band with a minimum at the pseudo-cubic R-point, just as it is expected from the folding of Z onto R. The weight of this band, however, is far lower than that of the main band at Z. This is not a matrix-element effect but reflects the small Fourier component of the orthorhombic ionic potential leading to a low initial state spectral weight of the back-folded band~\cite{voit2000}.
We calculate the spectral weight within DFT by 'unfolding' the orthorhombic unit cell bands into the primitive cubic unit cell, projecting the orthorhombic wave functions onto the cubic wave functions (red dots in Fig.~\ref{main3}~e))~\cite{Ku:2010,Popescu:2012}. This shows that spectral weights are expected to be highly concentrated on a single band with minimum at Z, fully consistent with our experiments. 
We note that the DFT calculation predicts small hybridization gaps between the main and back-folded bands. These gaps are comparable to the impurity scattering in our films and can thus not be resolved in our experiment.

For a quantitative discussion of our ARPES data, it is important to consider more subtle modifications of the band structure. The orthorhombic distortion splits the $d_{xy}$ orbital from the $d_{xz}$ and $d_{yz}$ orbitals 
and changes bond-angles and thus the hopping integrals. This leads to a significant lifting of degeneracies. For instance at the pseudocubic Z-point, we find 4 occupied bands split by $\approx 250$~meV (Fig.~3~e)), which is not negligible in comparison with electronic self-energies. This highlights the importance of a careful treatment of structural distortions for a discussion of correlation effects in \SMO{} and related distorted perovskites.

Obtaining precise atomic positions in oxide thin films from experiment is a daunting task. For our electronic structure calculations we thus use the experimental unit cell parameters and determine the full structure by relaxing the ionic coordinates within DFT.
As shown in Ref.~\cite{hampel2021} the structural properties of \SMO{} are highly sensitive to correlation effects. 
We take this into account by performing the structural relaxation within DFT+U with $U=3$~eV and $J=0.7$~eV. Using non-magnetic calculations performed with VASP, this results in structural parameters for bulk SMO close to the ones obtained with full DFT+DMFT calculations~\cite{hampel2021}. 
We note that a realistic uncertainty of the tilt angle of the MoO$_6$ octahedra of $\pm 1^{\circ}$ corresponds to a change in bare band width of $\approx 40$~meV or roughly 1\%. This is negligible for the below discussion of electronic self-energies.

% Experiment vs DMFT
%
% i) Extract self-energy by taking primary band as reference
% ii) Simulate spectral function by combining unfolding in DFT with DMFT self-energies
%
We now proceed to estimate electronic self-energies by taking the principal DFT band $\varepsilon^{\text{DFT}}(k)$ where the spectral weight is concentrated as a reference for the bare band.
The experimental quasiparticle dispersion $\varepsilon^{\text{QP}}(k)$ is obtained by fitting energy and momentum distribution curves (EDCs and MDCs) with Lorentzians and taking their peak positions. Assuming that the imaginary part $\Sigma''$ is constant over the width of the quasiparticle peak and neglecting low-weight bands over the range of interest, the real part of the self-energy is obtained from $\Sigma'(\varepsilon) = \varepsilon^{\text{QP}}(k) - \varepsilon^{\text{DFT}}(k)$. 
$\Sigma'$ derived in this way is shown in Fig.~\ref{main4}(a) and is in good agreement with our DFT+DMFT calculation performed with $U=3.40$~eV and $J=0.33$~eV for the relaxed atomic positions in the experimental unit cell~\footnote{Note that $U$ and $J$ here refer to the interaction parameters of the Kanamori Hamiltonian and differ from $U$ and $J$ used in DFT+U.}. 
We first note that $\Sigma'$ is nearly linear over an energy range of several hundred meV. 
This is in good agreement with our DMFT calculations and distinguishes \SMO{} clearly from strongly correlated systems such as ruthenates where the self-energy shows a much more pronounced curvature and kink-like features~\cite{mravlje2011,tamai2019}.
The structural distortions in our epitaxial \SMO{} films break the cubic symmetry and cause a splitting in the self-energy for the $d_{xy}$ and $d_{xz/yz}$ orbitals evident in our DMFT calculations. However, this splitting is small and can be neglected for a discussion of the physical properties of \SMO.

At low energy, we obtain a renormalization factor \mbox{$\lambda \equiv -\frac{d\Sigma'(\varepsilon)}{d\varepsilon}\big|_{\varepsilon = 0} \approx 0.7$} from experiment, corresponding to a quasiparticle residue $Z = 1/(1+\lambda) \approx 0.6$ in good agreement with $Z=0.58$ obtained from our DMFT calculations.
Using the experimental mass enhancement $m^*/m_b = Z^{-1}$ and the bare density of states from our DFT calculations, we estimate an electronic specific heat coefficient $\gamma^{\text{ARPES}} \approx 7.8$~mJ/(mol$\,$K$^2$), in excellent agreement with direct measurements of \mbox{$\gamma \approx 7.9$~mJ/(mol$\,$K$^2$)} in \SMO{} single crystals~\cite{ikeda2000, nagai2005, zhao2007, wadati2014}. This suggests that our experimental electronic structure data and self-energies are representative of bulk \SMO.

In Figure~\ref{main4}(b,c) we show a direct comparison of our ARPES data with DFT+DMFT spectral functions that incorporate the effect of structural distortions. To do so, we write the DMFT spectral function as a linear combination of all bands with weights obtained from the unfolding of the DFT bands. We further add a frequency independent broadening of $\Sigma''=-0.25$~eV to the DMFT self-energy to simulate the effect of impurity scattering in ARPES. Overlaying the experimental quasiparticle positions (red dots in Fig.~\ref{main4}(b,c)) on this theoretical spectral function shows an excellent overall agreement for the dispersion of the principal band. This is also evident in the comparison of experimental and theoretical EDCs in Fig.~4(c).
We note, that our valence band spectra (supplementary information, section III) show significant high-energy spectral weight forming a broad peak near $-2.5$~eV.
A similar feature was detected previously in hard x-ray photoemission measurements and was interpreted as a plasmon satellite~\cite{wadati2014}. The theoretical analysis of this peak is beyond the scope of our work, because we only treat low energy states within the DMFT study presented here. However, this peak was also not observed in GW+DMFT studies which should incorporate the necessary energy scales and underlying physics~\cite{nilsson2017,petocchi2020,zhu2021}.

The remarkably good overall agreement of our ARPES data and DFT+DMFT calculations establishes the basic electronic structure of \SMO{} and demonstrates that \SMO{} is a highly coherent metallic oxide with weak to moderate electronic correlations. We note that this is a non-trivial finding, especially when placed in the broader context of other oxides of the 4$d$ transition-metal series. Indeed, the bare band width of \SMO{} of $\approx 3.7$~eV (in the cubic room-temperature structure) is 
comparable to that of the Ruddelsden-Popper series of ruthenates. 
Furthermore, the Mo$^{4+}$ and Ru$^{4+}$ ions have a particle-hole symmetric configuration and very similar values of 
the Hubbard-Kanamori interaction parameters $U\simeq 3.4$~eV and $J\simeq 0.33$~eV. 
Yet, the strength of correlations as measured by the quasiparticle residue $Z$ is markedly different for molybdates and ruthenates. 
The present work establishes $Z=m_b / m^{*}\approx0.6$ for \SMO, roughly a factor of 2 above $Z\approx0.27$ for SrRuO$_3$ (as estimated from specific heat measurements~\cite{kikugawa2015,deng2016}) and approximately 3 times larger than the quasiparticle residue for the $xy$ orbital in Sr$_2$RuO$_4$~\cite{mackenzie2003,mravlje2011,tamai2019}. 

This observation highlights the importance of more subtle aspects of the bare band structure than just the bandwidth in 
controlling the strength of correlations. Indeed, in the ruthenate series, the Fermi level is close to a van Hove singularity and 
this low-energy feature leads to an enhancement of correlation effects, as demonstrated in both theoretical and 
experimental studies of Sr$_2$RuO$_4$~\cite{mackenzie2003,tamai2019}.
This is primarily a consequence of the band filling and is not related to the dimensionality. In the quasi two-dimensional molybdate Sr$_2$MoO$_4$, the $xy$-derived band is not close to a van Hove 
singularity, in contrast to its ruthenate analogue, and indeed a recent DMFT study predicts a much lower effective mass enhancement, comparable to that found here for SrMoO$_3$~\cite{karp2020}. 
%It is interesting however to note that in this material, the mass enhancement of the $xz$ and $yz$ derived bands is substantially higher than the one found here for SrMoO$_3$. This can be explained by the much smaller bandwidth of these bands in the layered 214 compound (stemming from their quasi one-dimensional character) as compared to the quasi-cubic 113 compound. 

Together with the reduced correlations in \SMO, we find an extended coherence scale. This manifests itself in the highly linear real part of the self-energy and the absence of a marked crossover from strongly renormalized low-energy quasiparticles to weakly renormalized incoherent high-energy excitations which is typical for correlated metals~\cite{byczuk2007,mravlje2011,georges2013,tamai2019}. Consistent with the behavior of the self-energy, the Fermi liquid regime in the resistivity of \SMO{} extends to comparatively high temperatures~\cite{nagai2005,cappelli2021}. 
This suggests weak electron-phonon coupling in \SMO, as it can also be deduced from the good agreement of the ARPES self-energy with our DMFT calculations that do not include electron-phonon coupling. The remarkable transport properties of clean \SMO{} are thus likely dominated by the electron-electron interactions quantified in our work.
Understanding transport in \SMO{} quantitatively on this basis remains an interesting challenge for theory.

% CONCLUSIONS
%
In conclusion, we presented a comprehensive ARPES data set from epitaxial \SMO{} thin films grown by PLD. We show that the ARPES data are in good agreement with DFT+DMFT calculations that incorporate the effect of structural distortions. Our work demonstrates that \SMO{} is a weakly correlated metal with highly coherent electronic states over a large energy range.

\begin{acknowledgements}
We thank J. Fowlie, M. Hadjimichael, C. Lichtensteiger, W. Rischau for discussions and help with some of the experiments. This work was supported the Swiss National Science Foundation (SNSF) grants 184998, 177006, 165791, 182695, the SNSF Ambizione fellowship 161327, the
Comunidad de Madrid (Atracción de Talento grant No. 2018-T1/IND-10521) and the Spanish Ministry of Science and Innovation (MICINN - PID2019-105238GA-I00, MICINN-FEDER RTI2018-097895-B-C43).
CED acknowledges support from the National Science Foundation under Grant No. DMR-1918455. 
%MR and EBG acknowledge support from the SNSF project 182695. (included above)
The Flatiron Institute is a division of the Simons Foundation.

\end{acknowledgements}

\bibliography{SMO.bib}

\end{document}